\def\year{2016}\relax
\documentclass[letterpaper]{article}
\usepackage{aaai16}
\usepackage{times}
\usepackage{helvet}
\usepackage{url}
\usepackage{courier}
\usepackage{amsfonts}
\usepackage{amsmath}
\usepackage{color}
\usepackage{graphicx}
\usepackage{tabularx}
 \usepackage{multirow}
\usepackage{paralist}
\begin{document}
\title{TGSum: Build Tweet Guided Multi-Document Summarization Dataset }
\author{
	Ziqiang Cao$^1$ ~~ ~~ Chengyao Chen$^1$ ~~ ~~ Wenjie Li$^1$ ~~ ~~ Sujian Li$^2$ ~~ ~~ Furu Wei$^3$  ~~ ~~ Ming Zhou$^3$\\
	$^1$Department of Computing, The Hong Kong Polytechnic University, Hong Kong\\
	$^2$Key Laboratory of Computational Linguistics, Peking University, MOE, China \\
	$^3$Microsoft Research, Beijing, China\\
	{\tt \{cszqcao, cscchen, cswjli\}@comp.polyu.edu.hk  } \\
	{\tt lisujian@pku.edu.cn} \\
	{\tt \{furu, mingzhou\}@microsoft.com} \\
}
\maketitle
\begin{abstract}
\begin{quote}
%The performance of learning-based summarization models has been significantly hampered by the limitation of available human summaries.
%Manual Annotation requires extensive efforts. 
%This paper proposes an effective way to automatically collect large scales of news multi-document summaries with reference to social media's reactions.
%We utilize two types of common social labels in tweets, i.e., hashtags and hyper-links.
%Hashtags are used to cluster documents into different topic sets.
%On the other hand, linked tweets, namely tweets with hyper-links, often act as highlights which provide certain key points of documents.
%However, a linked tweet often contains noise and is incomplete, inappropriate to become a summary. 
%We thereby ``synthesize'' more reliable reference summaries which are expected to cover most key points of tweets.
%We adopt the ROUGE metrics to measure the coverage ratio, and develop Integer Linear Programming formulas to discover the upper bound of ROUGE. 
%Since we allow summary sentences to be selected from both documents and high-quality tweets, the collected model summaries can be abstractive. 
%Both informativeness and readability of the collected summaries are verified by manual judgment. 
%In addition, we train a Support Vector Regression summarizer on DUC generic multi-document summarization benchmarks.  
%With the collected data as extra training resource, the performance of this summarizer improves a lot on all the test datasets. 
The development of summarization research has been significantly hampered by the costly acquisition of reference summaries.
This paper proposes an effective way to automatically collect large scales of news-related multi-document summaries with reference to social media's reactions.
We utilize two types of social labels in tweets, i.e., hashtags and hyper-links.
Hashtags are used to cluster documents into different topic sets.
Also, a tweet with a hyper-link often highlights certain key points of the corresponding document.
We synthesize a linked document cluster to form a reference summary which can cover most key points.
To this aim, we adopt the ROUGE metrics to measure the coverage ratio, and develop an Integer Linear Programming solution to discover the sentence set reaching the upper bound of ROUGE. 
Since we allow summary sentences to be selected from both documents and high-quality tweets, the generated reference summaries could be abstractive. 
Both informativeness and readability of the collected summaries are verified by manual judgment. 
In addition, we train a Support Vector Regression summarizer on DUC generic multi-document summarization benchmarks.  
With the collected data as extra training resource, the performance of the summarizer improves a lot on all the test sets. 
We release this dataset for further research\footnote{\url{http://www4.comp.polyu.edu.hk/~cszqcao/}}.
\end{quote}
\end{abstract}

\section{Introduction}
%The rapid growth of on-line digital content publishing and propagation calls for an efficient automatic summarization system.
%So far, the learning-based models have become the major summarization approaches.
%Despite decades of research, the quality of a generated summary is still far from satisfactory.
%A big bottleneck of supervised summarizers is the lack of human summaries.
%For instance, the generic multi-document summarization task means summarizing a cluster of documents telling the same topic.
%In this task, the most widely-used datasets are published by Document
%Understanding
%Conferences\footnote{\url{http://duc.nist.gov/}} (DUC) in 01, 02 and 04.
%In total, there are only 139 document clusters with 376 reference summaries.
%%On average, for one task in DUC/TAC, there are only about 50 document sets with 4 reference summaries in each. 
%The limited labeled data force a learning-based summarization system to heavily rely on well-designed features.
%Sometimes unsupervised approaches (e.g., \cite{rioux2014fear}) even outperform them.

The rapid growth of on-line digital content calls for efficient automatic summarization systems. 
So far, the learning-based models have become the dominant summarization approaches. 
Despite decades of research, the quality of a machine generated summary is still far from satisfactory. 
A big bottleneck of supervised summarizers is the lack of human summaries used for training. 
For instance, the generic multi-document summarization task aims to summarize a cluster of documents telling the same topic. 
In this task, the most widely-used datasets are published by Document Understanding Conferences\footnote{\url{http://duc.nist.gov/}} (DUC) in 01, 02 and 04. 
Totally, there are 139 document clusters with 376 human reference summaries. 
The limitation of labeled data forces a learning-based summarization system to heavily rely on well-designed features. 
Sometimes unsupervised approaches (e.g., \cite{rioux2014fear}) even outperform supervised ones.

%Basically, two major aspects restrict manual annotation.
%On one hand, summarization is an extremely time-consuming and labor-intensive task.
%Before writing a summary, annotators have to read and understand the whole document(s).
%On the other hand, this task is very subjective. 
%Even experts fail to reach a consensus.
%As a result, DUC has to provide multiple reference summaries for a relatively objective evaluation.

Basically, there are two major factors restricting manual annotation.
On the one hand, summarization is an extremely time consuming and labor-intensive process. 
Before writing a summary, annotators have to read and understand the whole document(s). 
On the other hand, it is very subjective. 
Even experts fail to reach a consensus. 
As a result, DUC has to provide multiple reference summaries in order to have a relatively objective evaluation. 
In this paper, we suggest an effective way to automatically collect large-scales of news multi-document summaries with reference to social media's reactions on Twitter. 
Previously, a series of NLP tasks have tried to utilize the social annotations like followers~\cite{chen2014inferring},  emoticons~\cite{zhao2012moodlens} and responses~\cite{hu2014convolutional} etc. 
Here two kinds of common social labels, i.e., hyper-links and hashtags are leveraged for our purpose. 
A hashtag in the news area often serves for the brief description of an event. 
For example, given ``\#BangkokBlast'', we know the tweets with this hashtag all talk about the recent terrorist attack in Bangkok. 
Therefore, it is a nice indicator to cluster documents into the same topic. 
On the other hand, we take advantage of linked tweets (i.e., tweets with hyper-links) to generate the optimal \textbf{reference summary} for a document cluster. 
From our observation, linked tweets hold the following properties:
%This paper proposes an effective way to automatically collect large-scales of news multi-document summaries with reference to social media's reactions on Twitter.
%A series of NLP tasks have tried to utilize social annotations, such as hashtags~\cite{efron2010hashtag},  emoticons~\cite{zhao2012moodlens} and responses~\cite{hu2014convolutional}. 
%Here we leverage two kinds of common social labels, i.e., hyper-links and hashtags.
%A hashtag in the news area is often the description of an event.
%For example, just seeing ``\#BangkokBlast'' we know the corresponding news is about the recent terrorist attack in Bangkok.
%Therefore, it is a nice indicator to cluster documents into the same topic.
%On the other hand, linked tweets usually reflect users' concerns about the news, providing a series of key points.
%Thus we take advantage of them to generate the optimal \textbf{reference summary} for a document cluster.
%From our observation, linked tweets hold the following properties:
\begin{itemize}
	\item A large proportion of linked tweets can highlight key points of related news;
	\item Tags such as ``\#'' and ``@'' frequently appears in a linked tweet, bringing a large amount of noise;
	\item Due to the length limitation, most tweets are shortened and describe only one aspect of related documents.
\end{itemize}
The noise and incompleteness hamper a linked tweet to directly become a reference summary.
Take the tweets in Table~\ref{TB:TweetExample} as an example.
%Concisely, we elide URLs in each tweet.
%This document describes how Greece's Crisis drowns the life of a sardine fisherman.
%Since the local fish industry owns the world first robotic sardine-processing line, Tweet 1 is interested in how it works.
%Tweet 2 and 3 both describe the key points of this paper, like ``Greece'', ``fish industry'', and ``crisis'', but Tweet 2 can act as a title while Tweet 3 tells author's writing experience.
%Notably, although all the three tweets are able to indicate saliency, none of them is appropriate to a summary.
%Tweet 1 is shortened, Tweet 2 is not a complete sentence, and Tweet 3 contains many tags.
This document describes how Greece's Crisis drowns the life of a sardine fisherman. 
Since the local fish industry owns the world first robotic sardine processing line, Tweet 1 is interested in how it works. 
Tweet 2 and 3 both describe the key points of this paper, like ``Greece'', ``fish industry'' and ``crisis'', but Tweet 2 can be used as the title while Tweet 3 tells author's writing experience. 
Although all the three tweets are able to indicate saliency, none of them is appropriate to be a part of the summary. 
Tweet 1 is shortened, Tweet 2 is not a complete sentence, and Tweet 3 contains many tags.

\begin{table}[htb]
	\centering
	\small
	\begin{tabularx}{\linewidth}{c|X}
		\hline
		\multirow{3}{*}{Tweet} & (1) Software determines which are sardines and ... how much they weigh                                        \\ \cline{2-2} 
		& (2) A fish tale: How \#GreeceCrisis drowned an industry and a way of life                                    \\ \cline{2-2} 
		& (3) .@georgikantchev, @movingpicturetv, @vaniabturner and I look inside Greece's fish industry and see crisis \\ \hline
		News                   & The Fisherman's Lament -- A Way of Life Drowned by Greece's Crisis (http://t.co/FXGTUY3IBq)               \\ \hline
	\end{tabularx}
	\caption{An example of the tweets linked to the same news. URLs in tweets are elided for short.}
	\label{TB:TweetExample}
\end{table}
	
Notably, sentences in the news documents are usually well-written. 
Therefore, we do not directly treat linked tweets as reference summaries.
Instead, we select sentences from both the document cluster and high-quality tweets to form a reference summary which can cover most key points in tweets. 
This practice has the following advantages. 
First, the summary is far more readable than tweets and meanwhile provides a complete description of a news document cluster. 
Second, the length of a reference summary is controllable. 
A similar idea is adopted by~\cite{filippova2013overcoming}. 
They use the highlight and the first news sentence to build the sentence compression pair. 
Also because the highlights are usually not the complete sentences, they utilize highlights to indicate which syntactical structures in the original sentences can be removed. 
%In TGSum, a reference summary is expected to cover most key points of tweets. 
In TGSum, we introduce the most widely-used summary evaluation metrics ROUGE~\cite{lin2004rouge} to measure the coverage ratio of key points, and develop an Integer Linear Programming (ILP) solution to discover the sentence set which can reach the ROUGE upper bound.

Within one month, TGSum collects 4658 linked tweets in overall. 
Table~\ref{TB:DUC_Dataset} lists the basic information of TGSum and compares it with DUC datasets.
On average, a document cluster in TGSum contains 23 tweets, ensuring to generate reference summaries in different generalization degrees. 
In terms of the cluster number, our dataset has already exceeded the scale of DUC datasets, and it is still growing every day. 
Once the reference summaries are generated, we conduct extensive experiments to verify the quality and effect of this dataset. 
About 30\% summary sentences come from linked tweets, indicating summaries in TGSum are abstractive in certain degree. 
Manual judgment demonstrates that the majority of these summaries are informative and readable. 
In addition, we train a Support Vector Regression (SVR) summarizer on DUC generic multi-document summarization benchmarks. 
With the collected TGSum dataset as the extra training resource, the performance of SVR summarizer improves a lot on all test sets. 
%On average, a document cluster in TGSum contains 23 tweets, ensuring the potential to generate reference summaries from different text granularities.
%Thus we can expect to generate a more complete reference summary.
%From the perspective of the cluster number, our dataset has already reached the scale of DUC datasets, and it is still growing every day.
%Thus it shows a promising expectation to gather a large-scale multi-document summarization dataset. 
\begin{table}[ht]
	\centering
	\small
	\begin{tabular}{l|llll}
		\hline
		Dataset & Cluster \# & Doc. \# & Sent. \# & Ref. \# \\ \hline
		DUC 01  & 30         & 309         & 10639        & 60           \\ \hline
		DUC 02  & 59         & 567         & 15188        & 116          \\ \hline
		DUC 04  & 50         & 500         & 13129        & 200          \\ \hline
		TGSum   & 204        & 1114         & 33968       & 4658(tweets)             \\ \hline
	\end{tabular}
	\caption{Statistics of the summarization datasets.}
	\label{TB:DUC_Dataset}
\end{table}

%After the reference generation, we conduct extensive experiments to verify the quality and effect of this dataset.
%About 30\% summary sentences come from linked tweets, which means summaries in TGSum are abstractive.
%Manual judgment demonstrates these summaries are usually informative and readable. 
%In addition, we train a Support Vector Regression summarizer on DUC generic multi-document summarization benchmarks.  
%With the collected data as extra training resource, the performance of this summarizer improves a lot on all the test datasets.

The contributions of this paper are listed as follows:
\begin{enumerate}
	\item We propose to collect multi-document news summaries with the help of social media's reactions;
	\item We develop an ILP solution to generate summaries reaching the upper bound of ROUGE;
	\item We publish this dataset for further research.
\end{enumerate}

\section{TGSum Construction}
%To our knowledge, the only available linked tweet dataset is provided by \cite{lloret2013towards}.
%However, this dataset is quite small, only containing 100 news documents. 
This section explains how we build TGSum, i.e., a multi-document summarization dataset guided by tweets. 
There are 4 main steps. 
The URL acquisition step catches proper news URLs.
These URLs are in turn used in the data collection step to extract the linked tweets and news documents.
Afterwards, news documents are clustered based on the hashtags embedded in tweets. 
Finally, for the news documents to be summarized, an Integer Linear Programming (ILP) solution is developed to generate reference summaries which cover as many key points provided in tweets as possible. Below is the detailed description of these steps.

%\begin{figure}
%	\centering
%	\includegraphics[width=0.9\linewidth]{../PIC/FlowChart}
%	\caption{Flow chart of the summarization data collection}
%	\label{fig:FlowChart}
%\end{figure}
 
\subsection{URL Acquisition}
At the beginning, we have tried to directly extract linked tweets through searching trends on Twitter. However, most trends are concerned with the entertainment circle, referring people to the picture or video pages. 
It is thereby not appropriate for document summarization. 
Thus, we design an alternative strategy which firstly discovers news URLs. 
Specifically, through the Twitter search function, 74 active news accounts like New York Times, Reuters and CNN who have published tweets within a month are selected as seed users. 
All the DUC news providers are included to ensure the generated dataset is uniform with DUC. 
Next, by applying the Twitter user streaming API, we track all seed users' tweets from August 13th to September 13th. 
Despite many replications of news titles, these tweets provide the URLs of hot news published by the corresponding news accounts. 
%These URLs are then regarded as the indexes to collect public tweets with the Twitter term search API. 
We do not focus on a particular domain or type of news. 
From observation, the collected pieces of news come from a wide range of topics, including finance, politics, sports, disaster and so on. 
This open-domain dataset gives us the chance to learn summarization behavior in different genres.
%This results in a corpus that not only news from different domains were contained, such as sports, science, technology, or culture

\subsection{Data Collection}
Given a news URL, we collect its document as well as linked tweets. 
The news content is retrieved with the open Python package newspaper\footnote{\url{https://pypi.python.org/pypi/newspaper}}. 
We just reserve the main body of a document. 
Then we apply the Twitter term search API to extract linked tweets, and conduct careful preprocessing as below:
\begin{compactitem}
	\item Discard retweets;
	\item Delete non-English tokens in a tweet;
	\item Remove tweets which contains less than 5 tokens;
	\item Merge identical tweets.
\end{compactitem}
At the end, we collect 13207 valid tweets from 4483 news documents.
%After these steps, there are XX linked tweets from XX different URLs reserved.
%Then the corresponding 
%A title is also treated as a reference summary.
%On average, a document is associated with 4 unique tweets.
%Although tweets for a single document may fail to cover all the key points.
%Therefore, it is likely that linked tweets cover all the key points of original documents.

\subsection{Cluster Formation}
%A single document may fail to describe the event completely.
It is common that different newswires publish multiple versions of documents about the same news. 
These documents are clustered together to present a more complete description of an event. 
With regard to our dataset, hashtags provide a simple and accurate way to achieve this goal. 
After removing general hashtags such as \#ThisWeek and \#ICYMI (i.e., In Case You Missed It), the hashtags related to news are usually the key phrases of the event. 
For instance, \#GreeceCrisis refers to the financial crisis of Greece. 
When we further restrict to cluster documents in the same day, it is very likely that the documents pointed by the same hashtag describe the identical news. 
For a document attached to no hashtag, we put it into an existing cluster if its TF cosine similarity with the cluster exceeds a threshold, e.g., 0.5 as we set. 
After removing clusters which have less than 3 documents or less than 8 linked tweets, we retain 1114 documents in 204 clusters.

\subsection{Reference Generation}
As mentioned above, most tweets are incomplete and contain noises, which are not suitably included in a summary directly. 
Since the sentences in the original news documents are usually well-written, we decide to ``synthesize'' reference summaries by selecting sentences from both news documents and high-quality tweets as long as they are good representatives of the news information. 
We expect the generated reference summaries could cover most key points of tweets. 
Here we adopt ROUGE to measure the coverage ratio. 
Through the analysis of ROUGE, we develop an ILP based solution to sentence selection, making sure the generated reference summary reaching the upper bound of ROUGE.
%The ILP formulas can be applied to any summary dataset for upper bound generation.
\subsubsection{Analysis of ROUGE Measurement}
%ROUGE has become a standard automatic evaluation metric for summarization since 2004.
ROUGE counts the overlapping units such as the n-grams, word sequences or word pairs between the two pieces of text. 
Take the widely-used ROUGE-2 as an example.
The coverage score between a candidate summary and a linked tweet $i$ is: 
% due to its high capability of evaluating automatic summarization systems~\cite{owczarzak2012assessment}.
%Its recall-oriented score is computed as follows:
\begin{equation} \label{eq:rouge}
	ROUGE - 2_i = \frac{{\sum\nolimits_b {Gain_i(b)} }}{{\sum\nolimits_b {{n_{twt_i}}(b)} }}
\end{equation}
where $b$ stands for a bi-gram, and $n_{twt_i}(b)$ is the number of bi-gram $b$ in the $i_{th}$ linked tweet.
$Gain_i(b)$ is the maximum number of bi-gram co-occurring in the candidate summary and linked tweet:
\begin{equation} \label{eq:gain}
	Gain_i(b) = \min \{ {n_{twt_i}}(b),{n_{cnd}}(b)\}
\end{equation}
Given a linked tweet, its total bi-gram count is a fixed value.
Therefore we can rewrite Eq.~\ref{eq:rouge} to
\begin{equation}
ROUGE - 2_i = {w_{b,twt_i}}\sum\nolimits_b {Gain_i(b)}
\end{equation}
where ${w_{b,twt_i}} = \frac{1}{{\sum\nolimits_b {{n_{twt_i}}(b)} }}$ could be regarded as the weight for the linked tweet.
For a set of linked tweets, ROUGE averages their scores. 
\begin{equation}
ROUGE - 2 = \frac{1}{K}\sum\nolimits_{i = 1}^K {({w_{twt_i}}\sum\nolimits_b {Gain_i(b))} }, 
\end{equation}

Use $z(s) \in \{0,1\}$ to indicate whether or not a sentence $s$ is selected in the candidate summary. 
We can represent the maximization of ROUGE-2 under the length constraint $L$ as the following optimization function:
\begin{align}
	\max \quad &\sum\nolimits_{i = 1}^K ({{w_{b,twt_i}}\sum\nolimits_b  \min \{ {n_{twt_i}}(b),{n_{cnd}}(b)\} } ) \\
	s.t. \quad &{n_{cnd}}(b) = \sum\nolimits_s {z(s) \times {n_s}(b)} \label{ct:count}\\
	&\sum\nolimits_s {z(s) \times |s|}  \le L \label{ct:summaryLength}\\
	& z(s) \in \{ 0,1\},\quad \forall s \label{ct:binary}
\end{align}
where ${n_s}(b)$ is the frequency of $b$ in sentence $s$. 
The meaning of each constraint as follows:
\begin{description}
	\item[Constraint~\ref{ct:count}] calculates the number of $b$ in the current summary.
	\item[Constraint~\ref{ct:summaryLength}] satisfies the summary length limit.
	\item [Constraint~\ref{ct:binary}] defines a binary variable  $z(s)$.  
\end{description}

All the above formulas are linear expect the gain function.
Note the minimization problem can be computed as follows:
\begin{align}
Gain_i(b) &\le {n_{twt_i}}(b) \label{eq:le1} \\
Gain_i(b) &\le {n_{cnd}}(b) \label{eq:le2} \\
Gain_i(b) = {n_{twt_i}}(b)\ &|| \ Gain_i(b) = {n_{cnd}}(b) \label{eq:or}
\end{align}
\textbf{Constraints}~\ref{eq:le1} and \ref{eq:le2} ensure $Gain_i(b) \le \min \{ {n_{twt_i}}(b),{n_{cnd}}(b)\}$, while \textbf{Constraint}~\ref{eq:or} can be solved by the big M formula or indicator constraints\footnote{\url{http://www-01.ibm.com/support/docview.wss?uid=swg21400084}}. 
But there is a much simpler solution in this task.
Since the objective function is maximization of $Gain_i(b)$, this constraint will be realized automatically.
Finally, the whole linear programming formulas are listed below:
\begin{align}
\max \quad &\sum\nolimits_{i = 1}^K ({{w_{b,twt_i}}\sum\nolimits_b {Gain_i(b)} } )  \label{eq:obj}\\
s.t.\quad &{n_{cnd}}(b) = \sum\nolimits_s {z(s) \times {n_s}(b)}  \notag \\
&\sum\nolimits_s {z(s) \times |s|}  \le L \notag \\
& z(s) \in \{ 0,1\}  \notag \\
& Gain_i(b) \le {n_{twt_i}}(b),\quad \forall b,i \notag \\
& Gain_i(b) \le {n_{cnd}}(b),\quad \forall b,i \notag
\end{align}

The work of \cite{li2013using} also proposes an linear programming function for the maximization of ROUGE.
Based on the fact that $min(a,x)=0.5(-|x-a|+x+a)$, they introduce auxiliary variables and convert the maximization of Eq.~\ref{eq:rouge} into
\begin{align*}
	\max \quad &\sum\nolimits_b {({n_{cnd}}(b) - {C_i}(b))}  \\
	s.t.\quad &{C_i}(b) \ge {n_{cnd}}(b) - {n_{twt_i}}(b),\quad \forall b,i \\
	&{C_i}(b) \ge {n_{twt_i}}(b) - {n_{cnd}}(b),\quad \forall b,i 
\end{align*}
where $C_i(b)$ is an auxiliary variable equal to $| {n_{cnd}}(b) - {n_{twt_i}}(b)|$ in the solution.
However, they do not provide details to tackle the case of multiple measurements.
Compared with their approach, our model represents the gain function without transformation, which greatly speeds up the solution procedure.
In the next section, we illustrate how to extend the model to multiple ROUGE variants. 

\subsubsection{Extension to Multiple ROUGE Variants}
The above-mentioned approach generates the optimal summary for ROUGE-2.
Actually, other ROUGE variants such as ROUGE-1 are also very useful~\cite{owczarzak2012assessment}.
Our preliminary experiments also show that only 49\% bi-grams in the linked tweets appear in the original documents, while over 83\% uni-grams can be found.
%Take the following sentence as an example:
%If we only consider ROUGE-2, the score of this sentence is 0.
%However, most of its words are actually used in reference summaries, indicating this sentence is still very important.
Thus we attempt to generate reference summaries that optimize both ROUGE-2 and ROUGE-1.
We modify Objective Function~\ref{eq:obj} into
\begin{align}
\max \quad &(1 - \lambda )\sum\nolimits_{i = 1}^K {({w_{b,twt_i}}\sum\nolimits_b {Gai{n_i}(b)} )} \notag \\
&+ \lambda \sum\nolimits_{i = 1}^K {({w_{u,twt_i}}\sum\nolimits_u {Gai{n_i}(u)} )} \label{eq:combinedObj},
\end{align}
where $u$ stands for a uni-gram and $\lambda \in [0,1]$ is the trade-off between ROUGE-2 and ROUGE-1.
When $\lambda$ comes close to $0$ (e.g., $0.0001$ in the experiment),  we implement the effect that the selected summary is the optimum of ROUGE-2 and its ROUGE-1 is as large as possible.

Although ILP is NP-hard in general, considerable researches have produced a number of effective solution tools. 
In this paper, we adopt the IBM CPLEX Optimizer\footnote{\url{http://www-01.ibm.com/software/commerce/optimization/cplex-optimizer/}} which is a high-performance mathematical programming solver.

%To obtain candidate summary sentences, we 
%Firstly, we use Stanford CoreNLP~\cite{manning-EtAl:2014:P14-5} to tokenize the text and split it into sentences.
%Then we use the text processing steps provided by ROUGE, including:
%\begin{itemize}
%	\item Stem words with Porter Stemmer\footnote{\url{http://tartarus.org/~martin/PorterStemmer/index.html}}.
%	\item Extract uni-grams and bi-grams with stopwords excluded.
%\end{itemize}
%Afterwards, 
To ensure the sentence-level readability, we choose declarative sentences in the documents and tweets as candidate summary sentences, and then adopt Eq.~\ref{eq:combinedObj} to generate reference summaries. 
In line with DUC, we set the summary length to 100 words. 

\section{Experiment}
%In this section, we firstly inspect the content of linked tweets.
%We check whether they really provide key points.
%Then we analyze the quality of the generated summaries.
%Both automatic and manual evaluations are conducted.
%Finally, we design a supervised summarization model to verify the effect of our dataset.

Firstly, we inspect the content of the collected linked tweets. 
We check whether they really provide certain news key points. 
Then, we analyze the quality of the generated summaries. 
Both automatic and manual evaluations are conducted. 
Finally, we design a supervised summarization model to verify the effect of the TGSum dataset.

\subsection{Linked Tweet Analysis}
During the data collection, we find that most linked tweets are simply the extraction of news titles, but still each document can receive three unique linked tweets on average. We categorize these unique tweets in accordance with the summarization task.

\begin{description}
	\item[Extraction:] The tweet is directly extracted from original text.
	\item[Compression:] The same word sequence in a tweet can be found in a document sentence (except extraction).
	\item[Abstraction:] Over 80\% words of a tweet can be found in the news document, and located in more than one sentence.
	\item[Other:] The rest tweets. From our observation, most of them are comments.
\end{description} 

\begin{figure}[th]
\centering
\includegraphics[width=0.7\linewidth]{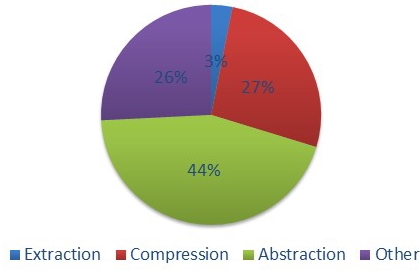}
\caption{Proportions of linked tweet types.}
\label{fig:PieChart}
\end{figure}

An example is provided for each tweet type in Table~\ref{TB:TypeExample}.
All the four tweets are from a news cluster describing “\#BangkokBlast”. We use bold font to indicate the words appearing in the news. 
It is noteworthy that the original document uses the word ``believe'' whereas the abstraction tweet chooses its synonym ``think''.
The second and third tweets are really salient and condensed, which demonstrates Twitter users are willing to summarize the documents. 
The proportions of different tweet types are shown in Fig.~\ref{fig:PieChart}. 
Note that our definition makes some abstraction tweets misclassified into the Other type. 
Thus the real abstraction tweets should occupy a larger proportion.
The direct extraction behavior (except title) in linked tweets seems to be rare. 
Users prefer adding tags to raise social communication. 
%For example, they will change “BBC” into “@BBC” or rewrite ``the crisis of Greece'' into ``\#GreeceCrisis''. 
In total, over 70\% of linked tweets are compression or abstraction, which shows its potential applications in learning the summary sentence generation models.
In the generated reference summaries, we find 30\% sentences come from linked tweets.
Therefore TGSum provides the abstractive version of references to some extent.

\begin{table*}[ht]
	\centering
	\small
	\begin{tabularx}{\linewidth}{c|XX}
		\hline
		Type        & Tweet                                                                                                                & Source                                                                                                                                                                                                                                                                                                                                                                                                                          \\ \hline
		Extraction  & Police have released a sketch of the main suspect                                                                    & \textbf{Police have released a sketch of the main suspect}, a man in a yellow T-shirt who was filmed by security cameras leaving a backpack at the shrine.                                                                                                                                                                                                                                                                               \\ \hline
		Compression & Suspect in Bangkok bombing is ``an unnamed male foreigner,'' according to an arrest warrant issued by a Thai court.    & The chief \textbf{suspect in} the deadly \textbf{bombing} of \textbf{Bangkok}'s popular Erawan Shrine \textbf{is ``an unnamed male foreigner,'' according to an arrest warrant issued} Wednesday \textbf{by a Thai court. }                                                                                                                                                                                                                                                    \\ \hline
		Abstraction & Taxi driver who thinks he picked up Bangkok bombing suspect says man was calm, spoke unfamiliar language on a phone. & A Thai motorbike \textbf{taxi driver who believes he picked up the suspect} shortly after the blast also said he did not seem to be Thai. ... who \textbf{spoke an unfamiliar language on} his cell \textbf{phone} during the short ride ... ... he still appeared very \textbf{calm}...  \\ \hline
		Other(Comment)     & Hmm, this face looks a bit familiar...                                                                                          & NULL                                                                                                                                                                                                                                                                                                                                                                                                                            \\ \hline
	\end{tabularx}
	\caption{Examples of different linked tweet types.}
	\label{TB:TypeExample}
\end{table*}

\subsection{Quality of TGSum}
\subsubsection{ROUGE Evaluation}
Now we verify the accuracy of our ILP-based ROUGE upper bound generation algorithm.
We choose $\lambda \in {0,1,0.0001}$, which means selecting sentences according to ROUGE-2, ROUGE-1 and their combination.
We refer to the corresponding model summaries as UB-1, UB-2 and UB-Combination respectively. 
As a contrast, we design a baseline applying the greedy algorithm.
It iteratively adds the sentence bringing the maximal ROUGE gain into the summary. 
Likewise, the summaries generated according to ROUGE-1 and ROUGE-2 are called GA-1 and GA-2. 
The ROUGE scores measured by linked tweets are shown in Table~\ref{TB:ROUGE_Comparison}. 
Obviously, the ILP solution achieves much larger ROUGE scores than the greedy algorithm. 
Then we compare three types of summaries derived from different $\lambda$ values. 
It is noted that both UB-2 and UB-Combination reach the upper bound of ROUGE-2, but the ROUGE-1 score of UB-Combination is higher. 
We thereby include the summaries generated by UB-Combination in TGSum as reference summaries.

\begin{table}[htb]
	\small
	\centering
	\begin{tabular}{l|ll}
		\hline
		Method         & ROUGE-1 & ROUGE-2 \\ \hline
		UB-1           & \textbf{55.24}   & 29.87   \\ \hline
		UB-2           & 49.15   & \textbf{34.74}   \\ \hline
		UB-Combination & 50.64   & \textbf{34.74}   \\ \hline
		GA-1           & 48.29   & 26.47   \\ \hline
		GA-2           & 46.10   & 29.44   \\ \hline
	\end{tabular}
	\caption{ROUGE(\%) comparison.}
	\label{TB:ROUGE_Comparison}
\end{table}

\subsubsection{Manual Evaluation}
We manually evaluate the sentence quality in the generated reference summaries. 
Two metrics are used, i.e., informativeness and readability. 
Since TGSum serves for learning-based summarization models, we do not consider the \textit{coherence} of an entire summary. 
For each metrics, a summary sentence is classified into three levels, namely good, OK and bad. 
Linked tweets are also evaluated for comparison. 
The results are presented in Table~\ref{TB:ManualEvaluation}. 
As seen, most sentences in the reference summaries are both informative and readable. 
The former owes to the saliency of most linked tweets, while the candidate sentence selection strategy brings the nice readability. 
Since a generated reference is the ROUGE upper bound measured by the set of linked tweets, a small amount of noised tweets will be excluded. 
Thus the number of unimportant sentences in reference summaries is much smaller than that in linked tweets. 
A bad case we find is the cluster about ``\#BBCBizQuiz''. 
All the tweets are questions. 
As a result, the reference summary fails to locate the salient sentences. 
With regard to readability, a large part of linked tweets are merely regarded as OK because they are not the complete sentences at all. 
By contrast, sentences in reference summaries are usually formal and readable. 
Quite a few summary sentences marked as OK are due to the splitting errors. 
For instance, sometimes a section title is attached to a sentence. 
Meanwhile, a pronoun at the beginning of a sentence brings the co-reference ambiguity problem.

%Then inspect the generated reference summary, and we find both the compression and abstraction tweets exist in it.
%While the first tweet is also important, it is not easy to automatically judge it is a complete sentence (lack of period).
%As a result, the upper bound algorithm choose one of its paraphrase.
%Since none of the words in the last tweet appears in news, it is ignored in the summary generation process.

\begin{table}[ht]
	\centering
	\small
	\begin{tabular}{l|l|ll}
		\hline
		\multicolumn{2}{l|}{Metrics}            & TGSum & TWEET \\ \hline
		\multirow{3}{*}{Informativeness} & Good & 0.66      & 0.60  \\
		& OK   & 0.29      & 0.23  \\
		& Bad  & 0.05      & 0.17  \\ \hline
		\multirow{3}{*}{Readability}     & Good & 0.64      & 0.17  \\
		& OK   & 0.23      & 0.51  \\
		& Bad  & 0.13      & 0.32  \\ \hline
	\end{tabular}
	\caption{Manual evaluation of TGSum's quality.}
	\label{TB:ManualEvaluation}
\end{table}

\subsection{Effect of TGSum}

To verify the effect of TGSum, we examine whether it can be used to improve the performance of summarization systems on DUC benchmarks.
We design a supervised Support Vector Regression (SVR) summarizer ~\cite{li2007multi,cao-EtAl:2015:ACL-IJCNLP}. 
To be concrete, each sentence in the training set is scored by ROUGE-2. 
Then we extract the features such as TF (the averaged TF scores of the sentence), LENGTH (sentence length), and STOP-RATIO (the ratio of stopwords), and train SVR to measure the saliency of sentences. 
For testing, we follow the greedy algorithm~\cite{li2014query} to select salient sentences into a summary.
According to~\cite{cao-EtAl:2015:ACL-IJCNLP}, the SVR summarizer achieves competing performance against the best participates in DUC.  
It is a standard learning-based summarization model, which is enough to emphasize the effect of training data.
We train this summarizer on different datasets and test it on DUC. 
The ROUGE results are shown in Table~\ref{TB:DUC_Performance}. 
Here ``DUC'' stands for DUC datasets except the testing year, and ``TWT\_REF'' means the same documents as TGSum but directly use tweets as reference summaries. 
Seeing from this table, ROUGE scores always enjoy a considerable increase when adding TGSum as an extra training set. 
In addition, even only given the TGSum dataset for training, the summarizer can still achieve comparable performance, especially on DUC 02. 
Thus, the effect of TGSum references is similar to the manual annotations of DUC, although the former is generated according to tweets automatically. 
In comparison to TGSum, directly using tweets as references does not always improve the summarization performance. 
The noise in tweets may misguide the model learning. 
For example, when treating tweets as references, the feature STOP-RATIO holds extremely high positive weight.
This improper feature weight can be ascribed to the informal writing style on Twitter, and it obviously fails to match the informativeness requirement.

\begin{table}[ht]
	\centering
	\begin{tabular}{l|lll}
		\hline
		Test set                  & Training set  & ROUGE-1 & ROUGE-2 \\ \hline
		\multirow{5}{*}{01} & TWT\_REF     & 29.23       & 5.45    \\
		& TGSum        & 29.40       & 5.73    \\
		& DUC       & 29.78       & 6.01    \\
		& DUC+TWT\_REF & 30.13       & 6.08    \\
		& DUC+TGSum    & \textbf{30.32}       & \textbf{6.26}    \\ \hline
		\multirow{5}{*}{02} & TWT\_REF     & 31.10       & 6.34    \\
		& TGSum        & 31.71       & 6.73    \\
		& DUC       & 31.56       & 6.78    \\
		& DUC+TWT\_REF & 31.74       & 6.80    \\
		& DUC+TGSum    & \textbf{32.15}       & \textbf{6.89}    \\ \hline
		\multirow{5}{*}{04} & TWT\_REF     & 35.73       & 8.83    \\
		& TGSum        & 35.97       & 9.16    \\
		& DUC       & 36.18       & 9.34    \\
		& DUC+TWT\_REF & 36.19       & 9.23    \\
		& DUC+TGSum    & \textbf{36.64}       & \textbf{9.51}    \\ \hline
	\end{tabular}
	\caption{Summarization performance with different training data.}
	\label{TB:DUC_Performance}
\end{table}

\section{Related Work}
\subsection{Summarization with Twitter}
Most summarization work on Twitter try to directly summarize tweets in a given topic.
Due to the lack of reference summaries, most researchers have to use unsupervised methods.
\cite{sharifi2010summarizing} detected important phrases in tweets with a graph-based algorithm.
But soon, the authors~\cite{sharifi2010experiments} developed a simpler ``Hybrid TF-IDF'' method, which ranked tweet sentences using the TF-IDF scheme and produced even better results.
A more complicated work was reported by \cite{liu2011sxsw},  which relied on Integer Linear Programming
to extract sentences with most salient n-grams.
It is worth mentioning that this paper highlighted the use of documents linked to the tweet set.
The saliency was measured according to TF in the documents, and their experiments demonstrated that allowing summary sentences to be selected from both tweets and documents achieved the best performance.
Recently, some papers~\cite{yang2011social,wei2014utilizing} simultaneously conducted single-document and tweet summarization based on cross-media features.

The above researches focus on tweet summarization, where documents provide additional features.
There are a limited number of papers about utilizing tweets to collect summarization data. 
The only work we know is~\cite{lloret2013towards}, who treated linked tweets as reference summaries and attempted to apply extractive summarization techniques to generate them.
They found that linked tweets were informative but their writing quality was inferior to extracted sentences.   
This work is equivalent to our practice which synthesizes reference summaries on the basis of linked tweets.
Moreover, their dataset only has 100 English news documents and only suits single-document summarization.
In contrast, we build a far larger dataset and apply it to multi-document summarization.

\subsection{ILP for Summarization}
Integer Linear Programming has been widely applied in summarization because it can appropriately model item selection state.
\cite{mcdonald2007study} originally introduced ILP in this area.
He constructed summaries by maximizing the importance of the selected sentences and minimizing their pairwise similarity, which was the extension of a greedy approach called Maximum Marginal Relevance (MMR) \cite{carbonell1998use}.
Given $N$ sentences,  his model contains $O(N^2)$ binary variables.
Thus it is quite inefficient when searching the optimal solution.
Later, \cite{gillick2009scalable} proposed to treat summarization as concept coverage maximization, where redundancy was implicitly measured by benefiting from including each concept only once.
They used bi-grams as the concept representation. 
The same idea was followed by many researches~\cite{woodsend2012multiple,li2013using}. Recently, \cite{schluter-sogaard:2015:ACL-IJCNLP} reported that syntactic and semantic concepts might also be helpful, and some papers such as \cite{cao2015ranking} combined sentence and concept selection process.
With heuristic rules, ILP can apply to compress~\cite{gillick2009scalable,berg2011jointly} or even fuse~\cite{bing-EtAl:2015:ACL-IJCNLP}.

\section{Conclusion}
% This paper proposes an effective way to automatically collect large-scales of news multi-document summaries with reference to social media's reactions.
%We use hashtags to cluster documents into different topic sets, and then ``synthesize'' reference summaries which are able to cover most important key points of linked tweets in a cluster.
%To measure the coverage ratio, we adopt the ROUGE metrics and develop Integer Linear Programming formulas to seek its upper bound. 
%%Since we allow summary sentences to be selected from both documents and high-quality tweets, the collected model summaries can be abstractive. 
%Manual evaluation verify the informativeness and readability of our dataset.
%In addition, we train a Support Vector Regression summarizer on DUC generic multi-document summarization benchmarks.  
%With the collected data as extra training resource, the performance of this summarizer improves a lot on all the test datasets. 

This paper presents an effective way to automatically collect large-scales of news multi-document summaries with reference to social media's reactions. 
We use hashtags to cluster documents into different topic sets, and then ``synthesize'' reference summaries which are able to cover most important key points embedded in the linked tweets within the cluster. To measure the coverage ratio, we adopt ROUGE metrics and develop an ILP solution to discover its upper bound. Manual evaluation verifies the informativeness and readability of the collected reference summaries. 
In addition, we train a SVR summarizer on DUC generic multi-document summarization benchmarks. 
With the collected data as extra training resource, the performance of this summarizer improves significantly on all test sets.

%This work is focused on the generic multi-document summarization task.
%However, we believe our dataset can be used in many other scenarios.
%On one hand, the compression type of linked tweets is an ideal source for the task of sentence compression.
%On the other hand, we are interested in adapting our dataset to the updating summarization task, just through tracking the same hashtag on different dates. 
The current work focuses on generic multi-document summarization. However, we believe our dataset can be used in many other scenarios. On the one hand, the compression type of linked tweets is an ideal source for learning sentence compression. On the other hand, we are interested in adapting our dataset to update summarization by tracking the same hashtag published on different dates.

%We leave this as our feature work. 
\section{ Acknowledgments}
The work described in this paper was supported by the grants from the Research Grants Council of Hong Kong (PolyU 5202/12E and PolyU 152094/14E) and the National Natural Science Foundation of China (61272291, 61273278 and 61572049). The correspondence authors of this paper are Wenjie Li and Sujian Li.

\bibliographystyle{aaai}
\bibliography{paper}

\end{document}